\documentclass[twocolumn]{aa}

\usepackage{psfig}
\usepackage{rotating} 

\usepackage{graphicx}
\newcommand{\msun}{\ensuremath{M_{\odot}}}
\newcommand{\lsun}{\ensuremath{L_{\odot}}}
\newcommand{\zsun}{\ensuremath{\rm Z_{\odot}}}

\newcommand{\hb}{\ensuremath{{\rm H}\beta}}

\newcommand{\teff}{\ensuremath{T_{\rm eff}}}

\newcommand{\gapprox}{\,\rlap{\lower 2.5pt % > ungefaehr =
\hbox{$\sim$}}\raise 1.5pt\hbox{$>$}\,}
\newcommand{\lapprox}{\,\rlap{\lower 2.5pt % < ungefaehr =
\hbox{$\sim$}}\raise 1.5pt\hbox{$<$}\,}
\begin{document}
   \title{The dynamical mass of the young cluster W3 in NGC 7252:} 
   \subtitle{Heavy-Weight globular cluster or ultra compact dwarf galaxy ?}

     \author{C.~Maraston \inst{1}
	  \and
	  N.~Bastian \inst{2,6}
           \and 
 	  R.~P.~Saglia \inst{3}
            \and 
          Markus Kissler-Patig \inst{2} 
            \and 
         Fran\c cois Schweizer \inst{4} 
            \and 
         Paul Goudfrooij \inst{5} 
               }

   \offprints{C.~Maraston}

 \institute{Max-Planck-Institut f\"ur Extraterrestrische Physik,
              Giessenbachstra{\ss}e,
                D-85748 Garching b. M\"unchen, Germany\\
                email: maraston@mpe.mpg.de
		\and
     European Southern Observatory,
     Karl-Schwarzschild-Str.~2, 85748 Garching, Germany
	\and
      Universit\"ats-Sternwarte M\"unchen, Scheinerstr. 1, 
      D-81679 M\"unchen, Germany
        \and
      Carnegie Observatories, 813 Santa Barbara Str., Pasadena, CA 91101-1292, USA
        \and 
      Space Telescope Science Institute, 3700 San Martin Drive, Baltimore, MD21218
        \and 
      Astronomical Institute, Utrecht University Princetonplein 5, 3584 CC 
Utrecht, The Netherlands
  } 

   \date{Received; accepted}

   \abstract{We have determined the dynamical mass of the most
     luminous stellar cluster known to date, i.e. object W3 in the
     merger remnant galaxy NGC~7252. The dynamical mass is estimated
     from the velocity dispersion measured with the high-resolution
     spectrograph UVES on VLT.  Our result is the astonishingly high
     velocity dispersion of $\sigma=45 \pm 5~{\rm km/s}$. Combined
     with the large cluster size $R_{\rm eff}=17.5 \pm 1.8~{\rm pc}$,
     this translates into a dynamical virial mass for W3 of ~$ (8 \pm
     2) \; 10^7 \msun$. This mass is in excellent agreement with the
     value ($\sim 7.2\cdot10^7~\msun$) we previously estimated from
     the cluster luminosity ($M_{\rm V}=-16.2$) by means of stellar
     $M/L$~ratios predicted by Simple Stellar Population models (with
     a Salpeter IMF) and confirms the heavy-weight nature of this
     object. This results points out that the NGC~7252-type of mergers
     are able to form stellar systems with masses up to $\sim~10^8~\msun$.
     We find that W3, when evolved to $\sim$~10 Gyr, lies far from the
     typical Milky Way globular clusters, but appears to be also
     separated from $\omega$Cen in the Milky Way and G1 in M31, the
     most massive old stellar clusters of the Local Group, because it
     is too extended for a given mass, and from dwarf elliptical
     galaxies because it is much more compact for its mass. Instead
     the aged W3 is amazingly close to the compact objects named
     ultracompact dwarf galaxies (UCDGs) found in the Fornax cluster
     (Hilker et al.~1999; Drinkwater et al.~2000), and to a miniature
     version of the compact elliptical M32.  These objects start
     populating a previously deserted region of the fundamental plane.
     }
   \maketitle
%
%________________________________________________________________

\section{Introduction}

About a decade ago, HST observations (Holtzmann et al.~1992) confirmed
the early ground-based discovery (Schweizer~1982) that mergers and
merger remnant galaxies host luminous, compact objects. These are
believed to be candidate young globular clusters (GCs) on the basis of
their blue colours, high luminosities, and compactness as derived from
the small radii. They were predicted to form during the (violent)
interaction caused by the merger event (Schweizer~1987). The
properties of these star clusters, such as their number, masses, ages
and metallicities have attracted a number of studies (see review by
Schweizer~1998) since they impact on the success of the scenario in
which elliptical galaxies form via merging of spirals. Leaving aside
pros and cons of this view, comprehensively discussed in
Kissler-Patig~(2000), the formation of star clusters during mergers is
nowadays an accepted fact. Besides helping in exploring the connection
to the host galaxy, the properties of young stellar clusters aid in
understanding the poorly known process of star cluster
formation. Among the various properties, the present paper focuses on
the cluster mass. A striking feature of the star clusters detected in
galactic mergers and merger remnants are the high stellar masses
reached by some of the members, as derived using $M/L$~ratios from
stellar population models. For example, in the ``Antennae'' galaxies,
the most luminous objects have masses up to a few $10^{6}~\msun$
(e.g. Zhang \& Fall~1999). Such {\it luminous} masses are larger than
the {\it mean} mass of old globular clusters in the Milky Way ($M\sim
2\cdot 10^{5} \msun$, e.g. Harris~1991), but are comparable to the
mass of the most massive galactic GCs, like 47 Tuc ($\sim 1.3\cdot
10^{6}~\msun$, Meylan \& Mayor~1986) especially considering that
(some) cluster mass will be lost. The percentage of mass loss due to
stellar evolution between 30 Myr and 15 Gyr is only $\sim~20
\%$~(Maraston~1998), but dynamical processes such as evaporation and
tidal disruption are also expected to reduce masses of GCs. However
the amount is difficult to estimate, since it depends on the exact
orbit of the cluster, its number of particles and the potential of the
parent galaxy. Recent N-body simulations for potentials appropriate
for spiral galaxies (Baumgardt \& Makino~2003) indicate values from
$\sim ~50 \%$~of dynamical mass-loss, to a negligible amount of
dynamical mass loss when the cluster is populous enough ($N > 10^6$).
For the potential of an early-type system similar predictions are not
yet available.

This work focuses on the most extreme case of a superluminous star
cluster known to date, i.e. object W3 in NGC~7252. Its
luminosity-derived mass is in the range $\sim10^{7}-10^{8}
\msun$~(Schweizer \& Seitzer~1998; Maraston et al.~2001). These values
are more than ten times larger than those of the star clusters in the
``Antennae'', in spite of W3 being significantly older ($t\sim 300 -
500~{\rm Myr}$, Schweizer \& Seitzer~1998, Maraston et
al.~2001). However, the mass determination via population synthesis
models is affected by uncertainties in the age determination. As
stellar $M/L$~ratios increase with age due to the fading of the light,
an overestimation of the cluster age leads to an overestimation of its
mass (other effects like metallicity and initial mass function playing
a minor r\^ole).

Given the potential impact of such high masses on the formation of
stellar systems, in particular in violent interactions and extreme
environments, the light-derived masses have to be checked by comparing
them with the values determined dynamically. The first succesful
attempt in this direction was by Ho \& Filippenko~(1996) for the young
star clusters of NGC~1569 (see Section~5), and more recently by Mengel
et al.~(2002) for those in the ``Antennae''. Here we have employed the
spectrograph UVES on VLT to obtain a high-resolution optical spectrum
of W3 in order to measure its velocity dispersion, and hence its
virial mass. This paper reports on the (spectacular) results for
W3. We will devote a follow-up paper to present the results for the
other star clusters measured during the same observing program.
\section{Determination of velocity dispersion}
\label{sigma}
   \begin{figure} 
%\centering
   \includegraphics[width=0.5\textwidth]{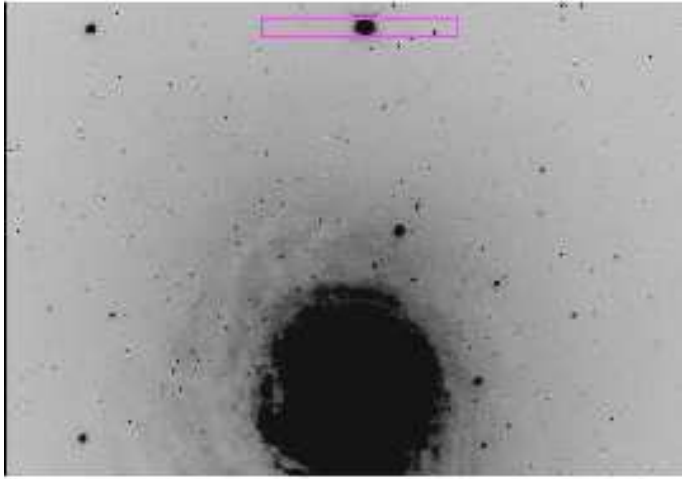}
   \caption {The portion of the WFPC2 image of NGC~7252
   containing W3 (from HST archive, see Miller et al.~1997), with the
   UVES slit position superimposed. The $x$~and $y$~dimensions are
   $27\farcs8$~and $18\farcs9$, or 8.7 and 5.9 kpc, respectively
   (1~pixel = 14.2 pc at the distance of NGC~7252 of 64.4 Mpc for
   $H_{0}=75~{\rm km/sec}$, see Miller et al.~1997).}
\label{slit}
   \end{figure}
We obtained a high-resolution, high signal-to-noise (S/N $\sim 25$~per
pixel), visual spectrum of W3 using the UltraViolet Echelle
Spectrograph (UVES) mounted on the ESO/VLT. The data were acquired
during 2001 in service mode. The spectrum covers the wavelength range
$3500-6600$~\AA~(setup Dichroic mode 1, CD\#2+\#3)) with a spectral
resolution of 2 km/sec. Standard stars of spectral types A to K were
acquired with the same setup in order to construct an appropriate
stellar template to be used to measure the velocity dispersion (see
below). The mean seeing during our runs was about 0.65 arcsecond FWHM
at 5000 Angstroems. A narrow slit ($0\farcs8$) was employed in
order to minimize the background contamination. The slit lenght was
$8\farcs$, the position angle was $15^{0}\pm~3^{0}$~and the slit was
centered on the object within less than 1 pixel, or
$0\farcs18$. Figure~\ref{slit} shows the portion of the WFPC2 image of
NGC~7252 that contains W3, with the UVES slit superimposed on the
object. The data were reduced by means of the UVES pipeline. The UVES
pipeline includes a standard procedure for the subtraction of the
background light, that applies well in case of a standard setting,
like that adopted by us. To strenghten our confidence concerning the
background contamination, we evaluate the surface brightness of the
unresolved light around W3 in a series of apertures around it from 15
to 20 pixels, and we found $V/arcsec^2=22.1 \pm~0.3$. Thus the galaxy
is much fainter than W3 ($V\sim17.8$) and the contamination is clearly
insignificant.
%
%----------------------------------------------------------- S_vib
%
%\begin{figure*}
%\begin{tabular}{ccc}
%\vbox{\psfig{file=../W3redlatotal4.ps,angle=270,width=5.5cm}} &
%%\vbox{\psfig{file=../W3redlb_total4.ps,angle=270,width=5.5cm}} &
%\vbox{\psfig{file=../W3redlc_total4.ps,angle=270,width=5.5cm}} \\
%\vbox{\psfig{file=../W3redld_total4.ps,angle=270,width=5.5cm}}&
%\vbox{\psfig{file=../W3redle_total4.ps,angle=270,width=5.5cm}} &
%\vbox{\psfig{file=../W3redlf_total4.ps,angle=270,width=5.5cm}}\\
%\vbox{\psfig{file=../W3redlg_total4.ps,angle=270,width=5.5cm}} &
%\vbox{\psfig{file=../W3redlh_total4.ps,angle=270,width=5.5cm}} &
%\vbox{\psfig{file=../W3redli_total4.ps,angle=270,width=5.5cm}} \\
%\vbox{\psfig{file=../W3redlj_total4.ps,angle=270,width=5.5cm}}&
%\vbox{\psfig{file=../W3redlk_total4.ps,angle=270,width=5.5cm}} &
%\vbox{\psfig{file=../W3redll_total4.ps,angle=270,width=5.5cm}}\\
%\vbox{\psfig{file=../W3redlm_total4.ps,angle=270,width=5.5cm}} &
%\vbox{\psfig{file=../W3redln_total4.ps,angle=270,width=5.5cm}}\\
%\end{tabular}
%      \caption {The individual echelle orders of the smoothed,
%	continuum subtracted and normalized spectrum of W3 (black
%	lines) as a function of $\ln \lambda $(\AA)~ (range in
%	\AA$\sim 5140-5570$). The red lines show the composite
%	template broadened to the fitted $\sigma$~(given in each panel
%	in km/sec.) and the green lines the difference with respect to
%	the smoothed spectrum of W3, shifted by 0.15 units for
%	clarity. The rms of the differences are given in each
%	panel. The whole $\lambda$-range is shown in the last panel.}
%	\label{spectrum}
%\end{figure*}
%
\begin{figure*}
 \includegraphics[width=\textwidth]{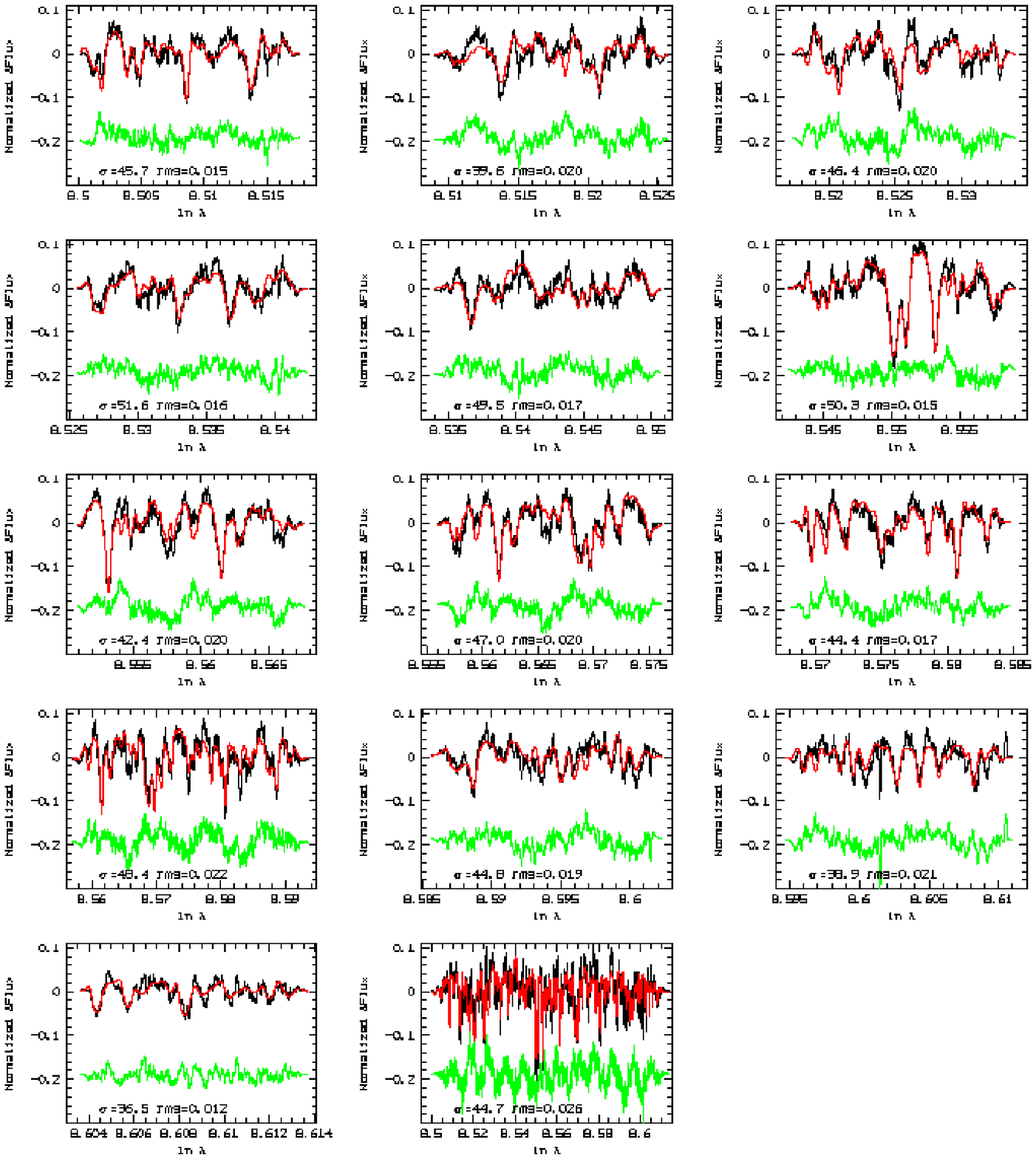}
      \caption {The individual echelle orders of the smoothed,
	continuum subtracted and normalized spectrum of W3 (black
	lines) as a function of $\ln \lambda $(\AA)~ (range in
	\AA$\sim 5140-5570$). The red lines show the composite
	template broadened to the fitted $\sigma$~(given in each panel
	in km/sec.) and the green lines the difference with respect to
	the smoothed spectrum of W3, shifted by 0.15 units for
	clarity. The rms of the differences are given in each
	panel. The whole $\lambda$-range is shown in the last panel.}
	\label{spectrum}
\end{figure*}
%______________________________________________________________

We determined the line-of-sight velocity dispersion of W3 using an
adapted version of the Fourier Correlation Quotient (FCQ, Bender 1990)
method as implemented by Bender, Saglia and Gerhard (1994). Following
Gerhard et al. (1998), the optimal order for the polynomial fitting of
the continuum was determined from Monte Carlo simulations, as well as
the statistical errors of the derived parameters. The wavelength range
was chosen to avoid intrinsically broadened lines (e.g., the Balmer
series), thus excluding the blue ($\lambda<4500$~\AA) section of the
UVES spectrum. In the reddest section ($\lambda>5650$~\AA) no strong
lines are available for kinematical analysis. The final results were
obtained using the region redwards of \hb~(see Figure 1, where the
spectrum of W3 has been blue-shifted to zero-redshift), comprising the
Mg and Fe lines.
%
%__________________________________________________ One column table
   \begin{table} \caption[]{Velocity dispersion $\sigma$~of W3 for
      different stellar templates (used wavelength range: $\sim$
      5064.44 $-$~$\sim$~5324.10). Statistical errors on $\sigma$'s are of the order of 5 km/s.} 
        \label{tabtempla}
      \begin{tabular}{ccc} \hline 
       Template & Sp & $\sigma$~(km/s) \\ 
      \noalign{\smallskip} 
       \hline
      \noalign{\smallskip} 
	HD 204943 & A7V & 37 \\
      \noalign{\smallskip} 
	HD 3229 & F5IV & 60.2 \\
      \noalign{\smallskip} 
	HD 8048 & F3V & 54 \\
      \noalign{\smallskip} 
	HD 8462 & F0V & 48 \\
      \noalign{\smallskip} 
	Composite & -- & 45 \\
      \noalign{\smallskip} 
	\hline 
	\end{tabular}
     \end{table}
%__________________________________________________________________
%
When spectra of stars as hot as the ones present in W3 are considered,
even for these metallic lines the intrinsic broadenings are much
larger than the UVES instrumental resolution ($\approx 37$ km/s for
the A7V star, see Table \ref{tabtempla}). Therefore, it is important
to test the effects of various templates. We considered: i) different
individual stellar templates; and ii) a composite stellar
template. The latter template aims at reproducing the photometric
properties of the cluster as close as possible and was obtained by
coadding individual stellar spectra with weights determined via the
best-fitting stellar population model for W3. Following Maraston et
al.~(2001), such a model is a Simple Stellar Population (SSP, i.e. a
single metallicity instantaneous burst) with age of 300 Myr and
metallicity $0.5~\zsun$, and it is shown to reproduce both the optical
and near-IR light extremely well. From this model we determine the
contributions to the visual spectrum (i.e. the region of application
of FCQ) of the various stellar temperature bins. These are:
$\sim~35\%$~for the main sequence stars around the turnoff
($\teff\ga8000~{\rm K}$), with other main sequence stars contributing
roughly $\sim 9\%$; $\sim~37\%$~for helium-burning giants ($4000 {\rm
K} \la \teff\la6000{\rm K}$); $\sim 19\%$ for Thermally Pulsing-AGB
stars. While template spectra for the TP-AGB stars are not available,
they are very cool and have small line widths, whence their inclusion
would {\it increase} the velocity dispersion (see Table~1).
The spectrum of the composite template is then: $\rm
S_{comp}=(0.5\cdot0.3681)\cdot(S_{\rm
HD3229}+S_{\rm HR8084})+(0.5\cdot0.09181)\cdot({S_{HR8462}}+{S_{HD204943}})+(0.35516\cdot S_{\rm HR3476})$.
We note that a similar template would have been obtained if we would
have considered for W3 an SSP with solar metallicity and the slightly
older age of $\sim~500~\rm Myr$, a model that also provides a very
good match to the optical spectrum of W3 (Schweizer \& Seitzer~1998).

The individual echelle orders of the continuum subtracted and
normalized spectrum of W3 (black lines) are shown in
Figure~\ref{spectrum}. For presentation reasons we smoothed the
original spectrum with a box average running mean of 20 pixel
radius. The heliocentric radial velocity derived averaging over the
echelle spectra presented in Figure~1 is $v_{\rm hel}=4822.5~\pm
0.97~{\rm km/sec}$, that compares very well with the value given by
Schweizer \& Seitzer~(1998) of $v_{\rm hel} = 4821~\pm 7~{\rm
km/sec}$.

We derived the value of the velocity dispersion $\sigma$
fitting each unsmoothed single order independently.  The red lines in
the Figure show the corresponding broadened composite template.  The
green lines display the difference to the smoothed spectrum of W3,
shifted by 0.15 units for clarity. The rms of the differences are also
given. The last panel shows the whole $\lambda$-range.

The $\sigma$'s derived from the single orders range from 36.5 to 51.6
km/sec, with a mean of 45 km/sec and r.m.s. of 5 km/sec. The r.m.s. of
the difference is in the range $1.5-2~\%$. Monte Carlo simulations
matching the S/N of the observations reveal no systematic bias. The
estimated (statistical) error for the single echelle order is 5
km/sec. For completeness, we list in Table~1 the $\sigma$'s obtained
for W3 using the individual stellar templates. As our final value of
the velocity dispersion $\sigma$~of W3 we adopt that obtained with the
composite template: $\sigma=45 \pm 5~{\rm km/sec}$. This $\sigma$~is
much larger than any value determined so far for galactic (see the
compilation by Pryor \& Meylan~1993) and extra-galactic (see
Sect.~5.1) GCs.
\section{Structure and radius of W3}
\label{mass}
We use the most recent version of Ishape (Larsen~1999) to measure the
structural parameters of W3 on the images obtained with HST/WFPC2
through the F555W filter with an exposure time of 60 s. and available
through the HST archive. First, we created a synthetic PSF using
TinyTim \footnote{version 5.0 as available at
http://www.stsci.edu/software/tinytim} at the position of W3 on the
chip. For the input spectrum of TinyTim we used a blackbody with of
6500~$K$~(having tested that blackbodies of other temperatures did not
affect the results). We then checked if Ishape could resolve any
point-like sources at the distance of NGC~7252 (64.4 Mpc, see
e.g. Schweizer \& Seitzer~1998), by running it on 10 isolated bright
sources at various locations on the PC, W3 included. For these sources
a King (1966) profile with a concentration index (i.e. the ratio of
the tidal radius to the core radius) $c$~of 30 has been assumed. The
returned values of the FWHM ranged between 0.17 and 0.71 pixels,
showing the presence of both extended and point-like sources on the
PC. We then ran Ishape on W3 assuming various analytical models. %
%__________________________________________________ One column table
   \begin{table}
      \caption[]{Structural parameters of W3 for various analytical models, $b$~and $a$~refer to the minor and major axis respectively.}
         \label{structure}
         \begin{tabular}{ccccc}
            \hline
            \noalign{\smallskip}
Model  & FWHM (pix) & $R_{\rm eff}(pix/pc)$ & $b/a$ & $\chi^2_{\rm \nu}$  \\
            \noalign{\smallskip}
            \hline
            \noalign{\smallskip}
            King5 & 1.61 & 1.14/16.2 & 0.71 & 2.75 \\
            \noalign{\smallskip}
            King30 & 0.71 & 1.05/14.9 & 0.76& 1.2 \\
            \noalign{\smallskip}
            King100 & 0.48  & 1.23/17.5 & 0.79 & 1.0 \\
            \noalign{\smallskip}
            Moffat15 & 1.13 & 1.18/16.8  & 0.77 & 1.45 \\
            \noalign{\smallskip}
            Moffat25 & 1.69 & 1.15/16.3 & 0.73 & 2.3 \\
            \noalign{\smallskip}
            Gaussian & 2.28 & 1.14/16.2 & 0.75 & 4.02 \\
            \hline
         \end{tabular}
   \end{table} 
The output structural parameters for these models are given in
Table~\ref{structure}, where the number after `King' refers to the
concentration index, that after `Moffat' to the power index multiplied
by 10. The reduced $\chi^2$~for each model ($\chi^2_{\rm \nu}$),
normalized to the $\chi^2$~ of model King100 are given in the last
column. The expected statistical uncertainty is $\delta \chi^2_{\rm
\nu}=0.17$~and refers to 70 degrees of freedom.

As the more extended King models reproduce the cluster light profile
significantly better, it is very likely that W3 is an extended
object. Further, since the ratio of minor to major axis ($b/a$) is
found to be $\sim 0.75$~consistently for all models, it is safe to say
that W3 is significantly flattened. The effect of the fitting radius
on the derived cluster size has been assessed by varying it between 5
and 20 pixels, and the variations are of the order of 1 pc, therefore
well within the observational errors.  The derived effective radii for
the best fitting model (King100) range between 14.9 pc and 18.5 pc,
for all possible combinations of model profile and fitting
radius. From all fits, we conclude that the best value of $R_{\rm
eff}$~for W3 is 17.5 $\pm~1.8$~pc. This value is larger than those for
most young stellar clusters. For comparison, for the clusters in the
``Antennae'' typical effective radii are found to be $\lapprox~10~{\rm
pc}$ (Whitmore et al.~1999).

The $R_{\rm eff}$~we obtain here for W3 is significantly larger than
the 7 pc obtained by Miller et al.~(1997). This is likely due to the
extended envelope of W3, which fell beyond the small apertures used by
Miller et al.~(1997) for their size estimates. The aperture method
used missed flux and oversubtracted background (which contained
cluster-envelope light), with the final effect of underestimating the
cluster radius. To test this hyphotesis we ran Ishape on the other
clusters of NGC~7252 with extended wings, i.e. W6, W26 and W30, and on
those that do not have extended wings, i.e.  W32 and W35. For the
former we find sizes systematically larger than those given by Miller
et al.~(1997), while for the latter we derive values perfectly
consistent with those by Miller et al..
\section{Dynamical mass of W3}
\label{mass}
According to Spitzer~(1987), the virial mass $M$~of a cluster is
$M=3a{{\sigma_{\rm x}^2}r_{\rm h}/G}$
where $\sigma_{\rm x}$~is the one-dimensional velocity dispersion, the
factor 3 relating it with the actual velocity dispersion; $r_{\rm
h}$~($=1.3~R_{\rm eff}$, Spitzer~1987) is the half mass radius;
$a\sim2.5$~is the factor connecting the half-mass radius to the
gravitational radius, the latter being the radius to be used into the
virial theorem. Using the value of $\sigma$~obtained with the
composite template, and $R_{\rm eff}=17.5~\pm 1.8~{\rm pc}$, we
obtain: $M_{\rm W3}=8 \pm 2\cdot 10^7~\msun$.
As will be discussed in Section~5, this value is in excellent
agreement with that we derived previously using stellar $M/L$ ratios
predicted by evolutionary population synthesis models (see Maraston et
al.~2001).
The assumption behind the determination of a virial mass is that W3 is
a relaxed system, i.e. self-gravitating and stable.

The half-mass relaxation time according to Spitzer~(1987) is $t_{\rm
rh}~={1.7\cdot10^{5}\cdot {r^{3/2}_{\rm h}}\cdot
N^{1/2}}\cdot{m^{-1/2}}$
where $r_{\rm h}$~is the half mass radius (in pc), $N$~is the total
number of stars, $m$~is the typical stellar mass (in $\msun$),
i.e. $m=M/N$, with $M$~the total mass of the cluster. $N$~is
determined by integrating the Salpeter mass function from
100~$\msun$~down to a lower limit of $0.1~\msun$, and normalizing the
IMF scale factor to the present mass taking into account both living
stars and stellar remnants (according to the prescriptions given in
Maraston~1998). The number of stars is estimated to be $N\sim2.8\cdot
10^8$. Using $r_{\rm h}=22.75~{\rm pc}$~(see Section~3), we obtain
$t_{\rm rh}\sim 6\times 10^{11}~{\rm yr}$. 
We conclude that, contrary to usual globular clusters, W3 is not
collisionally relaxed. Nevertheless, its crossing time is very short
($t_{cross}\approx R_{\rm eff}/\sigma=3.5\times10^5$~yr), much smaller
than its age ($\sim 300$~Myr, see above). Therefore we expect that,
like in elliptical galaxies, violent relaxation has taken place,
virializing the system.

It is extremely unlikely that W3 is a superposition of smaller
clusters for the following reasons: 1) the surface brightness profile
of W3 is very smooth (almost a PSF); 2) W3 is located in a quite
uncrowded area $\sim 5~{\rm Kpc}$~from the center (see Miller et
al. 1997, Figure 6; Maraston et al. 2001, Figure 1) where not many
objects are found; and 3) the component clusters would need to possess
almost identical spectral type and radial velocity to be invisible in
the spectrum.
\section{Discussion}
\subsection{Mass}

\label{mass}
Our main result is the perfect agreement between the dynamical mass
determined here for W3, and the luminous mass as derived for it from
model $M/L$~ratios (Maraston et al.~2001).  Recalling those results, a
luminous mass $7.2\times10^{7}~\msun$~was obtained for a Salpeter IMF
with lower and upper mass cutoffs of 0.1 and 100~\msun, respectively;
$4.\times10^{7}~\msun$~for a Gould IMF (Gould, Bahcall \& Flynn~1997),
i.e. an IMF the slope of which flattens with respect to Salpeter at
the low mass-end ($M\leq0.6~\msun$). The mass derived using a Salpeter
IMF is amazingly close to the dynamical estimate. In case one prefers
the Gould-type IMF, roughly 45 \%~of the total mass of W3 inside the
half-mass radius should be dark, but it is clearly not possible to
discriminate among the two options. It should be noted that the same
agreement is found for the luminous masses as derived from the near-IR
M/L~(Maraston et al.~2001).

The modeling of W3 as a simple stellar population of 300 Myr,
$0.5~\zsun$~appears very self-consistent from every side, colours,
Balmer lines (Schweizer \& Seitzer~1998), $M/L$. This favours the idea
that W3 is a globular cluster, since all but one globular clusters in
the Milky Way are simple stellar populations. However, as initially
pointed out by Schweizer \& Seitzer~(1998), the mass of the evolved W3
will still be $>100$~times larger than the mass of typical galactic
GCs, having taken into account the stellar mass losses. Note that a
flat IMF all over the mass range, as sometime advocated in the
literature (e.g. Mengel et al.~2002) does not help in reducing
significantly the total mass of the aged W3. Quantitatively, the
decrease in the stellar mass between 300 Myr and 10 Gyr for a flat IMF
with exponent 1.5 (in the notation in which the Salpeter's one is
2.35) is only a factor 1.64 larger than in case of Salpeter or Gould
IMF (Maraston~1998).

Dynamical mass-loss, e.g. tidal stripping and evaporation, act in
removing mass. However evaporation is not important for such a massive
object, and we estimated that it can remove only 1\% of mass (see
Maraston et al. 2001). Tidal stripping might be more efficient, but as
stated in Section~1, its impact is difficult to assess for our object
since current N-body simulations (Baumgardt \& Makino~2003) refer to
clusters in spiral-like gravitational potential for the parent galaxy,
while NGC~7252 has a de Vaucouleurs profile (Schweizer~1982). However,
according to these simulations, the amount of dynamical mass-loss
decreases strongly with an increasing number of particles, therefore
it is probably negligible for W3 because of its very large number of
stars (cfr. Section 4).
%                                                One column figure
%----------------------------------------------------------- S_vib
   \begin{figure*} \centering
   \includegraphics[width=0.7\textwidth]{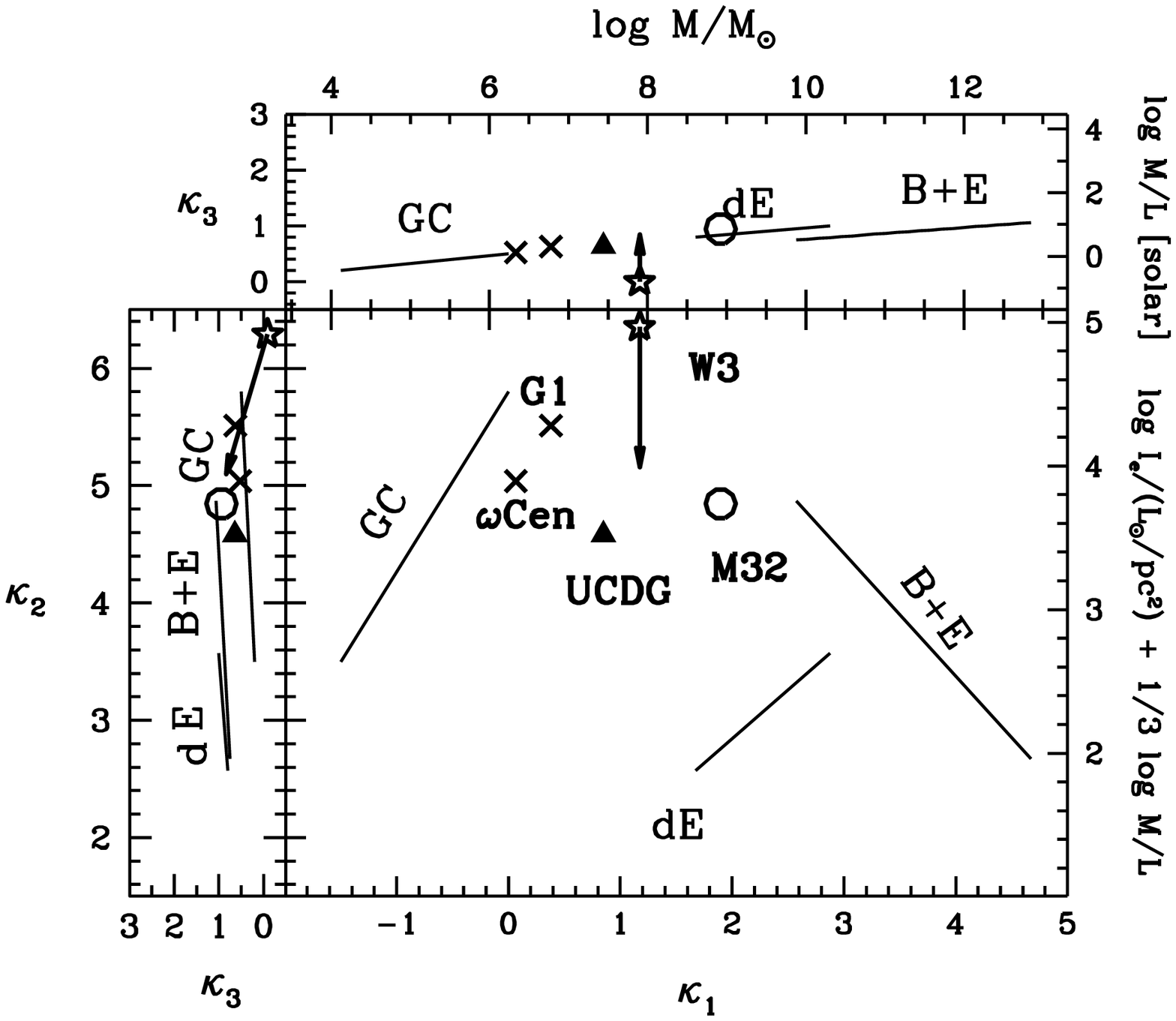}
   \caption {The $\kappa$~space of stellar systems, in a remake of
   Fig.~11 of Burstein et al.~(1997). To easy the reading, the
   physical parameters mass, $M/L$~and surface brightness are also
   reported, according to the definition of $\kappa$'s (Bender,
   Burstein \& Faber~1992). Lines define average sequences for: B+E
   (bulges and ellipticals); dE (dwarf ellipticals); GC (globular
   clusters). The open circle highlights the position of the compact
   elliptical M32. For the sources of these data see Burstein et
   al.~(1997). The quantities refer to $H_{0}=75~{\rm km/sec}$. The
   present position of W3 is displayed by means of the open star,
   while its predicted position at an age of 10 Gyr is indicated by
   the arrow. Crosses show G1 and $\omega$Cen, the most massive
   stellar clusters known in the Local Group. The filled triangle
   indicates the very compact objects detected in the Fornax cluster
   (Hilker et al.~1999; Drinkwater et al.~2000) named ultra compact
   dwarf galaxies (UCDGs) (sources of the data of the latter three
   objects are given in the text).}  \label{kappafig}
   \end{figure*}
%______________________________________________________________

The very large mass and flattened structure of W3 (cf. Table~2) cast
some doubt as to its nature as a GC. Indeed, in the Milky Way the
maximum value of $\sigma$~is $\sim 18~{\rm km/sec}$~for the old
metal-rich GCs NGC~6441 and NGC 6388 (Pryor \& Meylan~1993). The
Magellanic Clouds GCs, many of which have ages/metallicities like W3,
have normal luminosities ($\sim~10^{4 - 5}~\lsun$) and, when measured,
normal $\sigma$'s~(for the SMC GC NGC~419 Dubath et al.~(1997) give
$\sigma \sim~9.5\pm0.3~{\rm km/s}$). The young (some tenths of Myr)
star clusters of the nearby dwarf galaxy NGC~1569 have dynamical
masses $\sim 3\cdot 10^{5}~\msun$, and will evolve in perfectly normal
Galactic-type GCs. Instead, the mass of some of the star clusters in
the ongoing merger NGC~4038/4039 (``the Antennae''), although being
coeval to those of NGC~1569, are at least ten times larger (see
Section~1), as well as that of a very young ($\sim~15~{\rm Myr}$) star
cluster in the nearby spiral NGC~6946 (Larsen et al.~(2001).  Worth
noting is also the case of the star clusters in the 3 Gyr old merger
remnant NGC~1316 in Fornax. As discussed by Goudfrooij et al.~(2001),
the luminosity-derived mass of the brightest star cluster, being
already 3 Gyr old is still $\sim~1.4\cdot 10^{7}~\msun$. It would be
interesting to verify this number with dynamical measurements.

It seems natural to conclude that the formation of abnormally massive
star clusters is very much favoured by extreme environments like
galaxy interactions/mergers. 

With a mass of $\sim~10^8~\msun$~W3 is two orders of magnitudes more
massive than any other GC with reliable dynamical mass estimates, and
its classification does not appear straightforward.
\subsection{Clues from the $\kappa$~space}
\label{kappa}
In order to put the properties of W3 in context among other types of
stellar systems, the most comprehensive way is to consider the
fundamental plane of dynamically hot systems (Dressler et al.~1987;
Djorgovski et al.~1987) in which the basic structural parameters:
effective radius, mean surface brightness inside the effective radius,
and central velocity dispersion, are considered {\it
simultaneously}. Particularly useful is the re-definition of the
fundamental plane known as {$\kappa$}-space (Bender, Burstein \&
Faber~1992) that combines the three variables mentioned above into
more physically meaningful ones. The new variables called $\kappa_1$,
$\kappa_2$ and $\kappa_3$, are proportional to the mass, the product
of $M/L$~and surface brightness (i.e. the compactness of the system)
and the $M/L$, respectively.

The {$\kappa$}-space is shown in Figure~\ref{kappafig}, in a fashion
similar to Fig.~11 of Burstein et al.~(1997). The sequences defined by
the stellar systems: B+E (bulges plus ellipticals); dE (dwarf
ellipticals); GC (globular clusters) have been
drawn from their work.  The $\kappa$-coordinates of the compact dwarf
elliptical M32 (open circle) are from Bender et al. (1992).

In order to place W3 onto the $\kappa$-space, the central velocity
dispersion $\sigma_{0}$~has been evaluated from its average value
determined in this work, by means of the relation $\sigma_0 =
<\sigma>/0.87=51.72~\rm km/sec$~(derived from Djorgovski et al.~1997),
where it has been assumed that all the light out to $3-5 R_{\rm
eff}$~fell into the UVES aperture. The arrow indicates the position of
W3 when 10 Gyr old.  This implies a fading of 3.66 mag in B, according
to the evolutionary synthesis models adopted here (Maraston~1998;
Maraston et al.~2001).

The most massive star clusters (with dynamical mass measurements)
known before W3 are G1 in M31, the most luminous stellar cluster of
the Local Group, and $\omega$Centauri in the Milky Way. As the aged W3
could be more related to these heavy-weight objects than to typical
GCs, we have also plotted them into the $\kappa$-space. For G1 a
$\sigma_{0}\sim 27.8~{\rm km/sec}$~is provided by Meylan et
al.~(2001), and from their dynamical model we estimate $R_{\rm
eff}\sim 4.47~{\rm pc}$. For $\omega$Cen $\sigma_{0}\sim 16~{\rm
km/sec}$~(from Pryor \& Meylan 1993), and $R_{\rm eff}\sim 4.85~{\rm
pc}$~is from van den Bergh et al.~(1991).  It should be noted that
both G1 and $\omega$Cen have been argued to be the remnant nuclei of
stripped galaxies (Meylan et al.~2001; Freeman 1993), because they
show a metallicity spread, in other words they both are not simple
stellar populations as all other galactic GCs.  

The filled triangle in Figure~2 shows the position of the objects
named Ultra Compact Dwarf galaxies (UCDGs) detected recently in a deep
spectroscopic survey in Fornax (Hilker et al.~1999; Drinkwater et
al.~2000). The nature of these objects is currently under debate, and
as discussed by Phillipps et al.~(2001) they could be either extremely
large star clusters or extremely compact small galaxies, perhaps the
nuclei of stripped dE (see Drinkwater et al.~2003). This ambiguity
pushes to a comparison with our object. To place the UCDGs in the
$\kappa$-space we evaluate as average magnitude $B\sim~19.95$~using
objects n. 1,2,4,5 from Table~3 of Karik et al.~(2003) that show
consistent properties, i.e. we excluded object n.~3 that is
significantly brighter. Mean values of $R_{\rm eff}\sim 20.75~\rm
pc$~and $\sigma_{0}\sim~27.54~\rm km/sec$~are obtained by averaging
the individual values of the 4 objects (M.~Hilker, {\it private
communications}, see also Drinkwater et al.~2003). The spectra of the
UCDGs are consistent with old stellar populations (Hilker et al.~1999;
Drinkwater et al.~2003), therefore they should be compared to W3 aged
to 10 Gyr.

The following conclusions emerge from Figure~\ref{kappafig}. It is
confirmed that W3 lies far from the GC sequence because it is not
compact enough given its mass compared to the GC sequence. This
remains true when W3 is compared with G1 and $\omega$Cen. W3 is also
far from the position of dwarf ellipticals because it is far more
concentrated at a mass matching the least massive dEs. Instead, the
aged W3 shows properties that are remarkably similar to those of the
mysterious UCDGs. Using the similarity in the opposite direction
appears safer, since we presume to know how W3 has been formed,
i.e. during the merger event. This suggests that the UCDGs might have
formed in similar violent galaxy interactions at high redshift, since
they are found to be old.  It should be noted however that while these
objects are currently found in a galaxy cluster, NGC~7252 is rather
isolated. This implies that the cluster environment is not an
exclusive one for forming ultracompact massive objects. It would be
interesting to search for these type of objects in the vicinity of
other relatively isolated elliptical galaxies. Additionally, we note
that other gas rich galaxy mergers, like the Antennae, seem not to
have produced objects as massive as W3.

Finally, we briefly comment on the relation between W3 and the compact
elliptical M32 (empty circle). Similar to W3, M32 has properties that
place it in the realm of rare objects (see e.g. Bender, Burstein \&
Faber~1992), since for its mass is much more compact that the
elliptical galaxies defining the sequence in Figure~2. In this context
it is worth noting that, contrary to the compact objects of Fornax,
M32 is the only close companion of the giant spiral M31. Similarly, in
10 Gyrs W3 will be the only visible close companion of NGC~7252.

It is unfortunate that the most straightforward difference between GCs
and dwarf galaxies, i.e. the {\it simple} (i.e. single, mono-metallic
burst) versus the {\it complex} (i.e. extended star formation) nature
of their stellar populations is not assessable in case of distant
objects, like W3 (and the UCDGs). In any case, we have learned that the
galaxy interaction that has originated NGC~7252 roughly 1 Gyr ago was
also capable to form a stellar system whose structural properties fall
into a domain of rare objects. These objects at a given mass are more
compact than dwarf galaxies, and at given compactness are more massive
than GCs, and start populating a previously empty region of the
fundamental plane.

It would be valuable to explore under which conditions current
simulations of NGC~7252-type mergers (e.g. Mihos \& Hernquist~1996)
can produce W3-like objects.
\begin{acknowledgements}
CM and RPS thank Ralf Bender and Andi Burkert for very stimulating
discussions. We are also grateful to S{\o}ren Larsen for having
provided us with the most recent version of his code, and to him,
Daniel Thomas, Ulrich Hopp and Bo Milvang-Jensen for useful
comments. CM was supported at early stages of this work by the {\it
Deut\-sche For\-schungs\-ge\-mein\-schaft, SFB375}.\ FS gratefully
acknowledges partial support from the National Science Foundation
through grant AST-0205994.
\end{acknowledgements}

\end{document}